# Fixed Point Realization of Iterative LR-Aided Soft MIMO Decoding Algorithm

**Mehnaz Rahman**  *mehnaz@tamu.edu*
*Department of ECE*
*Texas A&M University*
*College Station, Tx- 77840*

**Gwan S. Choi**  *gchoi@ece.tamu.edu*
*Department of ECE*
*Texas A&M University*
*College Station, Tx- 77840*

**Abstract**

Multiple-input multiple-output (MIMO) systems have been widely acclaimed in order to provide high data rates. Recently Lattice Reduction (LR) aided detectors have been proposed to achieve near Maximum Likelihood (ML) performance with low complexity. In this paper, we develop the fixed point design of an iterative soft decision based LR-aided K-best decoder, which reduces the complexity of existing sphere decoder. A simulation based word-length optimization is presented for physical implementation of the K-best decoder. Simulations show that the fixed point result of 16 bit precision can keep bit error rate (BER) degradation within 0.3 dB for $8 \times 8$ MIMO systems with different modulation schemes.

## 1. INTRODUCTION

With the evaluation of wireless communication, multiple-input multiple-output (MIMO) systems have been adopted by different wireless standards such as IEEE 802.11n, IEEE 802.16e in order to achieve high data rates. Most of these standards have a specified minimum bit error rate (BER) or packet error rate (PER) to guarantee quality of service (QoS). Such as $10^{-6}$ is specified as maximum tolerable BER according to IEEE 802.11n standard [1].

The main challenge of MIMO system is to maintain the performance of the receiver with reduced complexity. Several algorithms have been proposed so far to address the issue offering different tradeoffs between performance and power consumption. The maximum likelihood (ML) detector offers optimal performance through exhaustive search, although its complexity increases exponentially with the number of transmitting and receiving antenna and bits in modulation [2, 3]. In contrast, linear detectors (LD) and sub-optimal detectors (zero forcing (ZF), minimum mean square error (MMSE) detectors) have been developed with significant performance loss.

Recently, lattice reduction (LR) has been proposed to achieve high performance with less complexity compared to the conventional K-best decoder [4, 5, 6]. These sub-optimal detectors are based on hard decisions. Hence, soft input-soft output (SISO) decoders are introduced with low density parity check (LDPC) decoder to achieve near Shannon performance with reasonable complexity [7, 8].

When considering practical implementation, fixed point design is a crucial step for hardware implementation in application specific integrated circuits (ASICs) or field programmable gate arrays (FPGAs). This paper presents a fixed point design of iterative soft decision based LR-aided K-best decoder proposed in [9], which exploits the algorithm of on-demand child expansion in iterative soft decoder. For soft decoding, the log likelihood ratio (LLR) values from the *K* best candidates are first computed for LDPC decoder and then, these are fed to the LLR update unit as inputs to the next iteration. This process of iteration is continued until the difference between the last two iterations becomes negligible. After that, the last updated LLR values are used for hard decision.

In this paper, we have conducted a novel study on fixed point realization of iterative LR-aided K-best decoder based on simulation. This process involves 2 steps: first is to select optimized architecture for each sub-module of K-best decoder, and the second one is to perform the fixed point conversion. The choice of proper architecture makes the hardware implementation easier, while the fixed point conversion minimizes the bit length of each variable. These objectives lead to the minimization of hardware cost, power, and area as well.

The simulation based optimization of word-length can minimize the total bit width of variables while obtaining similar BER. The results show that the total word length of only 16 bits can keep BER degradation within 0.3 dB for $8 \times 8$ MIMO with different modulation schemes. For QPSK modulation, precision of 16 bits results in less than 0.3 dB degradation, while 16 QAM and 64 QAM modulation provide 0.2 dB and 0.3 dB decrease in performance respectively compared to those of the floating bits of MIMO decoder.

The rest of the paper is organized as follow. In Section II we have introduced soft decision based LR-aided MIMO decoding algorithm. Next, in Section III fixed point realization of iterative decoder is proposed. Then, we analyze the results for all of our studied cases in Section IV. Finally, Section V concludes this paper with a brief overview.

## 2. SYSTEM MODEL

Let us consider a MIMO system operating in M-QAM modulation scheme with $N_T$ transmit antenna and $N_R$ receiving antenna. Hence, it can be represented as:

$$y^c = H^c s^c + n^c, \tag{1}$$

where $s^c = [s_1, s_2, \ldots s_{N_T}]^T$ is the $N_T$ dimensional transmitted complex vector, $H^c$ is complex channel matrix and $y^c = [y_1, y_2, \ldots y_{N_R}]^T$ is the $N_R$ dimensional received complex vector. Noise $n^c = [n_1, n_2, \ldots n_{N_R}]^T$ is a $N_R$ dimensional circularly symmetric complex zero-mean Gaussian noise vector with variance $\sigma^2$. The corresponding real signal mode is:

$$\begin{bmatrix} \Re[y^c] \\ \Im[y^c] \end{bmatrix} = \begin{bmatrix} \Re[H^c] & -\Im[H^c] \\ \Im[H^c] & \Re[H^c] \end{bmatrix} \begin{bmatrix} \Re[s^c] \\ \Im[s^c] \end{bmatrix} + \begin{bmatrix} \Re[n^c] \\ \Im[n^c] \end{bmatrix}$$

$$y = Hs + n, \tag{2}$$

where $s = [s_1, s_2, \ldots s_{2N_T}]^T$, $y = [y_1, y_2, \ldots y_{2N_R}]^T$ and $n = [n_1, n_2, \ldots n_{2N_R}]^T$. $\Re(\cdot)$ and $\Im(\cdot)$ denote the real and imaginary parts of a complex number respectively. ML detector solves for the transmitted signal by performing:

$$\hat{s} = arg_{\tilde{s} \in S^{2N_T}} \min \| y - H\tilde{s} \|^2 . \tag{3}$$

Here, $\| . \|$ denotes 2-norm, $\tilde{s}$ is the candidate vector, and $\hat{s}$ represents the estimated transmitted vector. This MIMO detection problem can be represented as the closest point problem in [10], and it performs a search through the set of all possible lattice points. In real signal mode, each antenna provides a search of 2 level: one for real part and the other for imaginary. The search is satisfied by the solutions with minimum error between sent and received signal. ML detector performs a search of all possible branches of the tree. Hence, it can obtain the maximum performance with exponentially increasing hardware complexity. Therefore, LR-aided detection is used to reduce the complexity of the ML detector [11]. Since lattice reduction requires unconstrained boundary, so the following change is made to (3) to obtain a relaxed search:

$$\hat{s} = arg_{\tilde{s} \in \mathcal{U}^{2N_T}} \min \|y - H\tilde{s}\|^2 , \qquad (4)$$

where $\mathcal{U}$ is unconstrained constellation set as $\{\ldots, -3, -1, 1, 3, \ldots\}$. But $\hat{s}$ may not be a valid constellation point, so a quantization step is applied:

$$\hat{s}^{NLD} = Q(\hat{s}), \qquad (5)$$

where $Q(.)$ is the symbol wise quantizer to the constellation set $S$. However, this type of naive lattice reduction (NLD) does not acquire good diversity multiplexing tradeoff (DMT). Hence, MMSE regularization is employed as follows:

$$\bar{H} = \begin{bmatrix} H \\ \sqrt{\frac{N_0}{2\sigma_2^2}} I_{2N_T} \end{bmatrix}, \qquad \bar{y} = \begin{bmatrix} y \\ 0_{2N_T \times 1} \end{bmatrix}, \qquad (6)$$

where $0_{2N_T \times 1}$ is a $2N_T \times 1$ zero matrix and $I_{2N_T}$ is a $2N_T \times 2N_T$ identity matrix [12, 13]. Eq. (4) can be represented as:

$$\hat{s} = arg_{\tilde{s} \in \mathcal{U}^{2N_T}} \min \|\bar{y} - \bar{H}\tilde{s}\|^2 . \qquad (7)$$

LR- aided detectors apply lattice reduction to the matrix $\bar{H}$ to find a more orthogonal matrix $\tilde{H} = \bar{H}T$, where $T$ is a unimodular matrix. This reduction effectively finds a better basis for the lattice and reduces the effect of noise and error propagation. After the reduction, the NLD with MMSE becomes

$$\hat{s} = 2T \arg\min_{\tilde{z} \in \mathbb{C}^{2N_T}} \left( \|\tilde{y} - \tilde{H}\tilde{z}\|^2 + 1_{2N_T \times 1} \right), \qquad (8)$$

where $\tilde{y} = (\bar{y} - \bar{H}1_{2N_T \times 1})/2$ is the complex received signal vector and $1_{2N_T \times 1}$ is a $2N_T \times 1$ one matrix. After shifting and scaling, (8) became the following one.

$$\hat{s} = 2T\tilde{z} + 1_{2N_T \times 1} . \qquad (9)$$

### 2.1 LR-aided Decoder
LR-aided K-Best decoder belongs to a breath first tree search algorithm. At a high algorithmic point of abstraction, the LR aided K-best search is performed sequentially, solving for the symbol at each antenna. In the beginning, QR decomposition on $\tilde{H} = QR$ is performed, where $Q$ is a

Mehnaz Rahman & Gwan S. Choi

$2(N_R + N_T) \times 2N_T$ orthonormal matrix and $R$ is a $2N_T \times 2N_T$ upper triangular matrix. Then (8) is reformulated as

$$\hat{s} = 2T \arg \min_{\tilde{z} \in \mathbb{Z}^{2N_T}} \left( \|\tilde{y} - R\tilde{z}\|^2 + 1_{2N_T \times 1} \right), \tag{10}$$

where $\tilde{y} = Q^T \tilde{y}$. The error at each step is measured by the partial Euclidean distance (PED), which is an accumulated error at a particular level for a given path through the tree. For each level, the *K* best nodes are calculated, and passed to the next level for consideration. At the last level, the *K* paths through the tree are evaluated to find *K* nodes with minimum PED for hard decision. In our adopted algorithm proposed in [9], the process of node calculation is optimized by on-demand child expansion.

### 2.2 On-demand Child Expansion
On-demand child expansion employs the principle of Schnorr-Euchner (SE) enumeration [14, 15] in a strictly non-decreasing order. It involves expanding of a node (child) if and only if all of its better siblings have already been expanded and chosen for future candidates [16].

Hence, at an arbitrary level of tree, the number of nodes needs to be expanded is bounded by $K + (K - 1)$ in the worst case scenario. For the entire tree, it becomes $4N_T K - 2N_T$. While considering the conventional K-best algorithm, the number of the expanded nodes is equal to $2N_T K m$, where m is the number of nodes of the modulation scheme. For instance, with list size, *K* = 4 and $N_T$ = 8, it requires 112 nodes for on-demand and 1024 nodes for conventional K-best algorithm to be expanded considering 16QAM modulation scheme. Therefore, use of on-demand child expansion requires significantly less computation, which reduces hardware complexity as well.

While working with soft decision, each path of chosen *K* best paths is considered as potential candidate. Hence, these K paths are passed to the soft-input soft-output (SISO) decoder for soft decoding.

### 2.3 Iterative Soft Decoding
LDPC decoder in [17] calculates approximate LLR from the list of possible candidates using (11).

$$L_E(x_k|Y) \approx \frac{1}{2} \max_{x \in X_{k,+1}} \left\{ -\frac{1}{\sigma^2} \| y - Hs \|^2 + x_{[k]}^T \cdot L_{A,[k]} \right\} - \frac{1}{2} \max_{x \in X_{k,-1}} \left\{ \frac{1}{\sigma^2} \| y - Hs \|^2 + x[k]^T \cdot LA,[k] \right\}, \tag{11}$$

where $x_{[k]}^T$ and $L_{A,[k]}$ are the candidates values {-1 or 1} and LLR values except k-th candidate respectively. In order to perform the soft decoding, the LLR values are first computed at the last layer of K-best search. Then, the soft values are fed into the iterative decoder for the subsequent iteration. This process is continued till the difference in error levels between the last two iterations becomes negligible. At the end, the last updated values are used for hard decision.

## 3. FIXED POINT REALIZATION

The system level diagram of adopted iterative LR-aided K-best decoder in [9] is presented in Fig. 1.

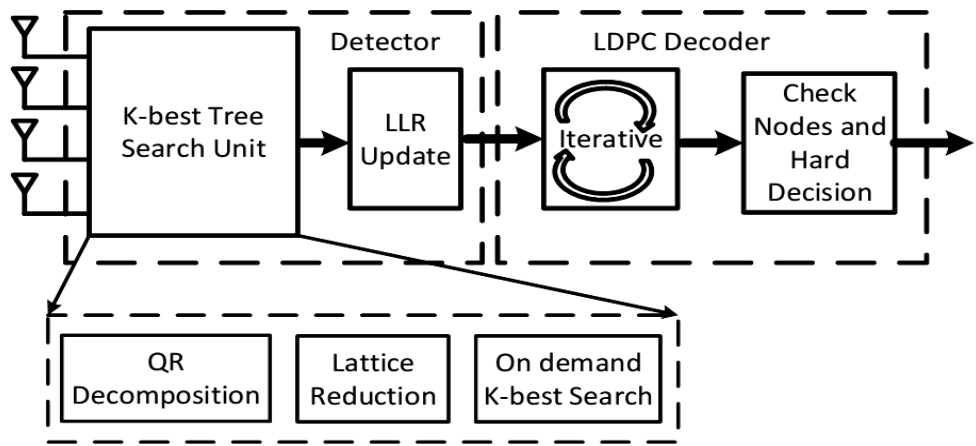

**FIGURE 1:** System level model of iterative LR-aided K-best decoder.

The fixed point realization of iterative LR-aided decoder involves two steps: First is the architecture selection of each sub-module and the second is the fixed point conversion. The selection of proper architecture makes the hardware implementation easier, and the fixed point conversion minimizes the bit widths of variables. Hence, it can reduce hardware cost including area, power and time delay.

### 3.1 Architecture Selection of Each Sub-module

#### 3.1.1 QR Decomposition
There are three well known algorithms for QR Decomposition proposed in [18]. Among them, the Givens rotation algorithm implemented by Coordinate Rotation Digital Computer (CORDIC) scheme under Triangular Systolic Array (TSA) in [19, 20] is selected for QR Decomposition. CORDIC is adopted due to its simple shift and operations for hardware implementation with reduced latency and it can be implemented easily exploiting parallel and pipeline architecture.

#### 3.1.2 Lattice Reduction
The Lenstra-Lenstra-Lovasz (LLL) algorithm proposed in [21] is a popular scheme for implementing lattice reduction. It can obtain optimal performance with low complexity. Hence, it is suitable for hardware realization by transforming the complicated division and the inverse root square operation into Newton-Raphson iteration and CORDIC algorithm respectively [22].

#### 3.1.2 LDPC Decoder
The hardware design of LDPC Decoder in [17] consists of separate LLR calculation unit. It takes one of the candidates at a given time and computes the LLR value at each clock cycle. Then, the new LLR is compared to the maximum of previous LLRs. Hence, this unit has to keep track of 2 values for each LLR. One for those whose k-th of the candidate list is 1 (Lambda-ML), and the other for 0 (Lamdba-ML-bar). After that, the LLR values are calculated as the subtraction of Lamdba-ML and Lamdba-ML-bar divided by 2.

### 3.2 Fixed Point Conversion With Word Length Optimization

In order to perform the fixed point conversion, all floating-point variable and arithmetic operations are converted into fixed point version. It is simulated by MATLAB HDLcoder, which is bit-accurate with verilog source code and mimics the actual operation in hardware.

Each word length is then optimized to determine the minimum bit width for each fixed point variable keeping high performance within tolerated error limit. To choose the length of proper precision bits, first minimum integer word length is calculated under large data simulation. After that, the minimum and maximum value of each variable is calculated through MATLAB profiling.

To estimate precision bits, first minimum and maximum fractional word length are chosen through extensive simulation. Then the bit error rate (BER) performances are evaluated for subsequently decreasing word length from max to selected min. At the end, the word length for which high performance with lower and tolerable error limit can be achieved, is selected as final optimized precision bit length.

## 4. SETUP AND RESULTS

This section demonstrates the performance of iterative soft decision based LR-aided K-best decoder in [9] for $8 \times 8$ MIMO with different modulation schemes. The signal to noise ratio (SNR) is defined as the ratio of received information bit energy to noise variance.

We first analyze the performance of four iterations of iterative LR-aided decoder with list size of 4 for different modulation schemes. We also demonstrate the comparison of performance for floating word length with that of fixed one. For iterative decoder, as shown in [9] the improvement gained from the 3rd to 4th iteration is limited and negligible for iteration beyond that. Hence, we consider BER versus SNR curve of 4th iteration in order to compare among maximum performances.

### 4.1 Simulation and Analysis

The performance of four iterations of $8 \times 8$ MIMO for QPSK modulation scheme is presented in Fig. 2.

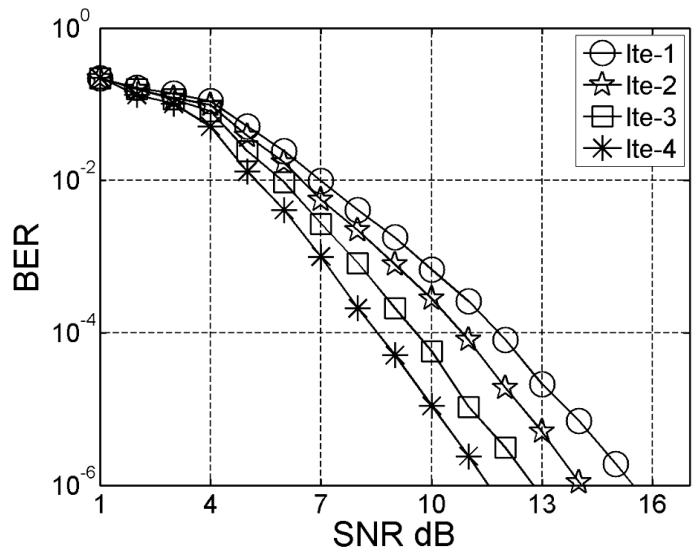

**FIGURE 2:** BER vs SNR curve of the first 4 iterations of 8 x 8 LR-aided decoder for QPSK modulation scheme with K as 4.

As shown in Fig. 2, for QPSK modulation scheme with list size of 4, we observe 1.5 dB improvement in BER due to the 2nd iteration at the BER of $10^{-6}$. When we compare the performance of 1st iteration with 3rd and 4th one, the improvement increases to 3.0 and 4.2 dB respectively. Next, the performances of four iterations for 16QAM and 64QAM modulation schemes are given in Fig. 3.

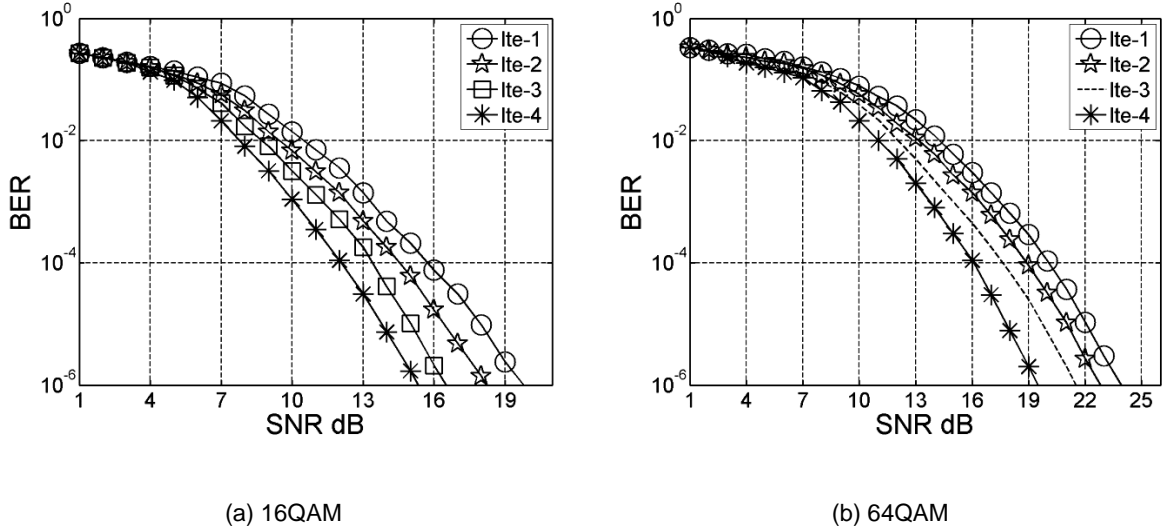

(a) 16QAM  (b) 64QAM

**FIGURE 3:** BER vs SNR curve of the first 4 iterations of 8 x 8 LR-aided decoder with K as 4.

As demonstrated in Fig. 3(a), the performance of 2nd iteration is approximately 1.5 dB better than the 1st one with K as 4 for 16QAM modulation scheme. When increasing the iteration, the performance improves by 3.0 dB for the 3rd and 4.0 dB for the 4th iteration compared to the 1st one.

For 64QAM modulation scheme having the same $K$ as in 16QAM, the improvement due to the 2nd iteration is 1.5 dB, as shown in Fig. 3(b). If we then compare the 3rd and 4th iteration with respect to the 1st one, the improvements are 3.0 dB and 4.0 dB respectively. Therefore, with iteration number, the performance between i-th and (i+1)-th iteration becomes saturated.

### 4.2 Comparison of Performance
The comparison of performance of iterative LR-aided decoder using floating bit length with that of fixed precision word length is presented in Fig. 4 for QPSK modulation scheme.

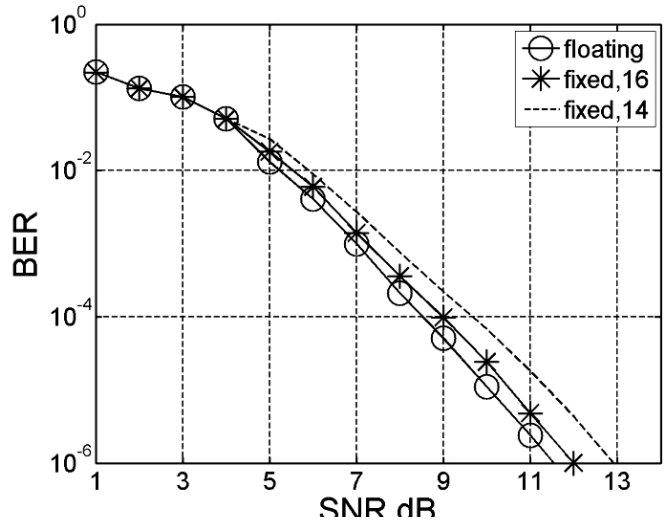

**FIGURE 4:** BER vs SNR curve of the 4th iteration of 8 x 8 LR-aided decoder for QPSK modulation scheme with floating and fixed word length of 14 and 16 bits.

The simulation is done for $8 \times 8$ MIMO system with *K* equal to 4. We consider only the 4th iteration in order to evaluate comparison among maximum performances. As shown in Fig. 4, when considering bit length of 16 bits, the performance degrades 0.3 dB comparing with the floating one. If we decrease the word length to 14 bits, the performance decreases to 1.3 dB. Hence, 16 bits of fixed word length can limit the performance degradation to 0.3 dB at the BER of $10^{-6}$.

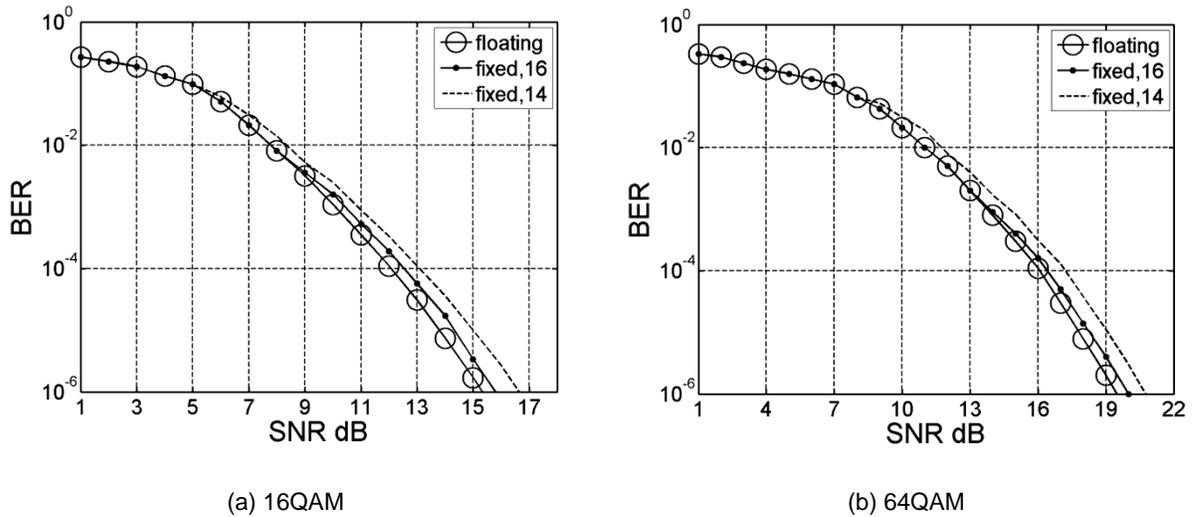

(a) 16QAM          (b) 64QAM

**FIGURE 5:** BER vs SNR curves of the 4th iteration of 8 x 8 LR-aided decoder with floating and fixed word lengths of 14 and 16 bits.

Next, Fig. 5 represents the performance curve of 4th iteration for 16QAM and 64QAM modulation scheme. As demonstrated in Fig. 5(a), for 16 QAM modulation scheme, 16 bit word length decreases the BER performance 0.2 dB at the BER of $10^{-6}$. When considering the word length of 14 bit, the performance degrades approximately about 1.3 dB.

While considering the performance of 64QAM, shown in Fig. 5(b), 16 bit precision limits the degradation to 0.3 dB. When evaluating for fixed 14 bits, the performance decreases to more than 1.3 dB. Therefore, 16 bits of fixed word length can keep the BER performance degradation within 0.3 dB for QPSK, 16QAM and 64QAM modulation schemes.

## 5. CONCLUSION

In this paper, we develop the fixed point design of an iterative soft decision based LR-aided K-best decoder. A simulation based word-length optimization provides feasible solution for hardware implementation with the selection of efficient architectural sub-components. Besides, the fixed point conversion also minimizes the bit width of each variable. Hence, it can reduce hardware cost including area, power and time delay.

Simulation results show that the total word length of only 16 bits can keep BER degradation about 0.3 dB for $8 \times 8$ MIMO with different modulation schemes. For QPSK modulation, precision of 16 bits results in less than 0.3 dB degradation, while 16 QAM and 64 QAM modulation provide 0.2 dB and 0.3 dB decrease in performance respectively compared to those of the floating bits of MIMO decoder. Future scope of this proposed work includes but not limited to the hardware implementation of LR-aided iterative K-best decoder in order to validate our fixed point design.